\documentclass[12pt,a4paper,reqno]{amsart}

\usepackage[margin = 1in]{geometry}
\usepackage{a4wide}
\usepackage{amssymb,amsmath,amsthm,mathrsfs}
\usepackage{graphicx}
\usepackage{caption}
\usepackage{enumerate}
\usepackage{upref}
\usepackage{framed} 
\theoremstyle{remark}
\newtheorem*{remark}{Remark}

\begin{document}
	\title[Sources and Sinks of Rare Trajectories Identified by Importance Sampling]{Sources and Sinks of Rare Trajectories in 2-Dimensional Velocity Fields Identified by Importance Sampling}
	\author{Meagan Carney*, Holger Kantz$\dagger$}
	\thanks{*Max Planck Institute for the Physics of Complex Systems,
		N\"othnitzer Str. 38, D 01187 Dresden, Germany. Email: meagan@pks.mpg.de\\ $\dagger$ Max Planck Institute for the Physics of Complex Systems, N\"othnitzer Str. 38, D 01187 Dresden, Germany. Email: kantz@pks.mpg.de.}
		\date{\today}
	\begin{abstract}
We use importance sampling in a redefined way to highlight and investigate rare events in the form of trajectories trapped inside a target coherent set. We take a transfer operator approach to finding these sets on a reconstructed 2-dimensional flow of the atmosphere from wind velocity fields provided by the Portable University Model of the Atmosphere. Motivated by extreme value theory, we consider an observable $\phi(x) = -\log(d(x,\gamma))$ maximized at the center $\gamma$ of a chosen target coherent set, where it is rare for a particle to transition. We illustrate that importance sampling maximizing this observable provides an enriched data set of trajectories that experience such a rare event. Backwards reconstruction of these trajectories provides valuable information on initial conditions and most likely paths a trajectory will take. With this information, we are able to obtain more accurate estimates of rare transition probabilities compared to those of standard integration techniques.
	\end{abstract}

\maketitle

\section*{}

\section{Introduction}

Atmospheric eddies play a major role in extreme weather phenomena such as heat-waves, hurricane movement, and pollution distribution \cite{RWB,pbook}. Topologically, these eddies can be seen as time-varying almost invariant sets (often referred to as coherent structures or sets) of a flow where there is minimal particle exchange across the boundary \cite{FSPD}. However, eddies are not the only coherent structures in the atmosphere. At mid-latitudes, the conservation of angular momentum guarantees the formation of another type: coherence as a naturally occurring consequence of particles trapped between the space of counter-rotating eddies. Due to time-delays, these structures often have longer lifetimes over a fixed region than those of a single eddy. Understanding where these coherent sets occur and the likelihood of trajectories ending up inside them can provide a new and useful perspective on atmospheric movement. 

We reconstruct a 2-dimensional model of atmospheric flow defined on a space $X\subseteq \mathbb{R}^2$ from wind velocity fields provided by the Portable University Model of the Atmosphere (PUMA) \cite{LBFJKLS}. Following recent literature, we estimate the transition probability matrix (TPM) of the flow by taking a fine grid of boxes and measuring transitions of particles from one box to another over a fixed time interval. We take a transfer operator approach by approximating the Perron-Frobenius operator with the associated TPM. Spectral properties of this operator provide information on the invariant (and almost-invariant) structure of the space. This approach is discussed in detail in \cite{DJ,F,FD,FP} and applied numerically to an ocean flow model in \cite{FSPD}. For adaptation purposes, our methods differ slightly from those of the listed literature by including variations on the TPM \cite{LK} and employing a spectral clustering approach equipped with $K$-means \cite{CAN,CK,V}; however the foundational arguments remain the same. 

Motivated by extreme value literature \cite{HN,evtbook}, we consider an observable 
\[
\phi(x) = -\log(d(x,\gamma))
\]
where $\phi: X\rightarrow\mathbb{R}$ is defined for every $x\in X$ and $\gamma$ is the euclidean center of a target coherent set. In this way, trajectories of the observable under the flow are maximized as they approach the center of the coherent set. Under some flow $f_t:X\rightarrow X$, it is often of interest to consider a set of random variables defined by $X_N = \phi\circ f_t(x_N)$ for a set of initial values $x_N\in X$ at some fixed time $t$. For our choice of $f_t$ and $\phi$, we show that the sequence $(X_N)$ for $N = 1,\dots, M$ behaves as though it comes from some unimodal distribution where $f_t(x_N)\rightarrow \gamma$ gives $X_N\rightarrow\infty$. This setup gives a natural correspondence between rare events occurring under the flow $f_t$ (e.g. it is rare for particles to transition into the almost-invariant set) and large values of $X_N$ where the distribution of $(X_N)$ decays in the tail.

We apply an importance sampling method, called genealogical particle analysis (GPA) \cite{D,WB}, that exponentially tilts the distribution of $(X_N)$ so that the probability of observing larger values (and hence, values of $f_t(x_N)$ closer to $\gamma$) is increased \cite{CKN,D,RWB, WB}. GPA works by killing and cloning trajectories under the flow at specified sampling times based on a weight function that determines the performance of a trajectory. Large values of the weight function indicate that a trajectory behaves as though it comes from the target (tilted) distribution. In the end, the surviving trajectories represent the set that has a higher probability of ending near $\gamma$. Backwards reconstruction of these trajectories allows us to find the set of most likely paths that end in the coherent set within a specified time interval.


We emphasize that the novelty of our method is not in the search for finite-time coherent sets in a flow (which has been widely studied in past literature, \cite{DJ,FD,FP,FSPD} to list a few); but in a new application of importance sampling algorithms \cite{CKN,WB,RWB} in this setting which allows us to obtain more accurate rare probability transition estimates into such a set.

\section{Description of Methods}
\subsection{Almost-Invariant Sets as Finite-Time Coherent Sets in the Portable University Model of the Atmosphere}

Coherent set estimation for nonautonomous systems has been studied extensively in the literature. Over finite time intervals, the transfer operator approach to search for almost-invariant sets in the autonomous case can be applied to find coherent sets in the nonautonomous case. We outline some main points below; however, for the interested reader we refer to \cite{DJ,F,FD,FP} containing some nice and detailed discussions. Let $Y\subset\mathbb{R}^2$ be compact and $V:Y\times \mathbb{R}\rightarrow \mathbb{R}^2$ be the smooth vector field on the domain $Y = [36^\circ, 70^\circ]\times[169^{\circ},205^{\circ}]$ (covering Europe) generated by the Portable University Model of the Atmosphere. Consider the nonautonomous ODE,
\begin{equation}\label{ODE}
\dot{x} = V(x,t)
\end{equation}
and $f_\tau:Y\times\mathbb{R}\rightarrow\mathbb{R}$ be the corresponding flow, e.g. $f_\tau(x_o,t_o)$ is a solution to (\ref{ODE}) with initial condition $x(t_o) = x_o$ with
\[
\frac{df_\tau}{dt}(x_o,t_o)|_{\tau=0} = V(x_o,t_o).
\]
Then $f_{t,\tau}(x)$ denotes the end position in $Y$ of a trajectory beginning at $x\in Y$, time $t$ and flowing for time $\tau$. A set $A\subset Y$ over the interval $[t,t+\tau]$ is called \textit{almost-invariant} if,
\[
\rho_{\mu,t,\tau}(A) = \frac{\mu(A\cap f_{t,-\tau}(A))}{\mu(A)}\approx 1
\]
where $f_{t,-\tau}(A)$ is the pullback of the set $A$ at time $t+\tau$ under $f$ by $\tau$ and $\mu$ is the (normalized) probability measure with the property that $\mu([a,b]\times [c,d])$ is equal to the area of a rectangular region $[a,b]\times[c,d]$ on $Y$. 

The Perron-Frobenius operator $\mathcal{P}_{t,\tau}: L^1(Y,m)\rightarrow L^1(Y,m)$ is defined by,
\[
\mathcal{P}_{t,\tau}(g(y)) = \frac{g(f_{t,-\tau}(y))}{|\det (Df_{t,-\tau}(y))|}
\]
where $m$ is the normalized Lebesgue measure and $g\in L^1(Y)$ is the normalized density function of $\mu$ \cite{FP}. 

Fixed points of $\mathcal{P}_{t,\tau}$ defined by $\mathcal{P}_{t,\tau}(g\cdot 1_A) = \lambda(g\cdot 1_A) = g\cdot 1_A$ indicate invariant sets on the space while almost-invariant sets correspond to values $\lambda\approx 1$. This is a consequence of the following nontrivial result from \cite[Prop. 5.7]{DJ},

\begin{equation}\label{graph}
\lambda \mu(A) = (\rho(A)+\rho(Y\setminus A)-1)\mu(A).
\end{equation}

When $\lambda\approx 1$ we have that the probability measure $\mu$ is close to the invariant measure of the system. In a similar way, if we consider the right hand side where the sets $A$ and $Y\setminus A$ form a partition of the space then finding an coherent measure can be viewed as a maximization problem of both $\rho(A)$ and $\rho(Y\setminus A)$. This approach is discussed in detail in \cite{FD}. Suppose there are $q$ almost-invariant sets on $Y$, then the extension of (\ref{graph}) is given by

\begin{equation}\label{graph2}
\max_{A_1,\dots, A_q} \rho(A_1,\dots,A_1) = \frac{1}{q}\sum_{k=1}^q\rho(A_k),
\end{equation}
by varying the partitions $A_1,\dots,A_q$ such that $A_k\cap A_\ell=\emptyset$ for $k\ne \ell$ and $\cup_{k=1}^q A_k = Y$.



Numerically, $\mathcal{P}_{t,\tau}$ is often approximated by a finite dimensional Galerkin approximation based on a fine partition $\{B_1,\dots,B_m\}$ of the space $Y$ \cite{DJ,FD, FSPD}. In this approach, the transition probability matrix that is formed under the flow from $t$ to $t+\tau$ is given by,
\[
P_{t,\tau,i,j} = \frac{m(B_i\cap f_{t,-\tau}(B_j))}{m(B_i)},
\]

where an entry $P_{t,\tau,i,j}$ is the probability of a uniformly selected point starting in box $B_i$ and ending in $B_j$. This discretization of the space turns the maximization problem described in (\ref{graph2}) into the graph theory equivalent min-cut problem. The sets that are found through the partition solution are exactly those coherent sets existing over the finite time interval $[t,t+\tau]$. 

\subsection{Importance Sampling (Finding Trajectories Likely to End in the Target Coherent Set)}\label{sec:IS}

We perform importance sampling on a larger domain $X = [30^{\circ}, 80^{\circ}]\times[160^{\circ},220^{\circ}]$ such that $Y\subset X$. Note that this does not change $f_{t,\tau}$ which is calculated over the whole domain $[-180^{\circ},180^{\circ}]\times [0^{\circ},360^{\circ}]$.

Let $A$ be a target coherent set on $Y\subset X$ estimated over the time interval $[t,t+\tau]$. We will require that $A$ be connected and define the center of $A$, $\gamma\in Y$ as the midpoint of $A$. We consider the observable 
\[
\phi(x) = -\log(d(x,\gamma))~~x\in X
\]
where $d$ is the Euclidean metric so that $\phi(x)$ is maximized as it approaches the center of the coherent set. 

Let $X_N = \phi\circ f_{t,\tau} (x_{N})$ for $x_N\in X$ be a sequence of random variables representing the value of our observable $\phi$ as a function of the end position of the $N$ trajectories on $X$ run under the flow $f_{t,\tau}$ from $t$ up to time $t+\tau$. For consistency of notation we will let $f_{t_n,t_{n+1}}(x_N)$ be the same flow starting at $(x_N)$ at time $t_n$ and ending at time $t_{n+1}$. 

Genealogical particle analysis (GPA) is an importance sampling method \cite{WB,D} that uses weights to perform a change of measure on the distribution of $(X_N)$ in a reversible way so that rare events are sampled more often. These weights can be thought of as measuring the performance of a trajectory at specified sampling times. Large values of the weight function imply that the trajectory $f_{t_n,t_{n+1}}(x_N)$ is behaving as if it comes from the target distribution. These trajectories will be cloned while low weight values indicate a trajectory that will be killed. Importance sampling algorithms are often used to lower relative error of tail probability estimation because the change of measure provides a set of trajectories that are more likely to end in a rare event. In our context, running GPA will provide a pool of trajectories that are \textit{most likely} to end in our coherent set over the time interval $[t,t+\tau]$.
%
%

One difficulty with GPA is determining a weight function that will change the measure in an appropriate way so that rare events are sampled more often. This choice depends on the distribution of $(X_N)$; however, it most commonly takes the form \cite{WB,RWB},
\begin{equation}\label{GPA}
\exp(V(\phi\circ f_{t_{n-1},t_n}(x_N))-V(\phi\circ f_{t_{n-2},t_{n-1}}(x_N)))
\end{equation}
which applies an exponential tilt by the function $V(x)$ to the distribution of $(X_N)$ at each sampling step $n$ where $t_n$ is divided evenly between the start time $t$ and the end time $t+\tau$. The weight function described by (\ref{GPA}) has proven numerically successful for importance sampling of random variables with symmetric, heavy-tailed distributions \cite{CKN,RWB,WB}. Following this guideline, we assume that $X_N$ is distributed according to a unimodal distribution with tails decaying to zero where an exponential tilt would result in larger sampling in the tail of the distribution of $X_N$. This assumption is supported numerically (see Figure \ref{fig:dist})). The explicit GPA procedure used for this analysis, where $V(x) = Cx$ and a trajectory is defined as a particles' movement through the atmosphere under the reconstructed PUMA flow, is described below.

\begin{framed}
	\textbf{The GPA Algorithm}
	
\begin{itemize}
	\item[1.] Initiate $N = 1,\dots,M$ starting particles uniformly distributed over the space $X$.
	\item[2.] For $n = 1\dots\lfloor \tau/\mathcal{T}\rfloor$ where $\tau$ is the total integration time. $\mathcal{T}$ is referred to as the \textit{resampling time}. 
	\begin{remark}
		It is important to balance $\mathcal{T}$ between the correlation time of $X_N$ and the Lyapunov time. Values of $\mathcal{T}$ taken too small can result in highly correlated trajectories (many of clones of a single trajectory) while too large can result in a relaxation back to the original distribution.
	\end{remark}
	\begin{itemize}
		\item[2a.] Iterate each trajectory from time $t_{n-1} = t+\mathcal{T}(n-1)$ to $t_{n} = t+\mathcal{T}n$.
		\item[2b.] At time $t_n$, stop the simulation and assign a weight to each trajectory given by,
		\[
		W_{N,n} = \frac{\exp(C (\phi\circ f_{t_{n-1},t_n}(x_{N})-\phi\circ f_{t_{n-2},t_{n-1}}(x_{N})))}{Z_n}
		\]
		where
		\[
		Z_n = \frac{1}{N}\sum_{N = 1}^M W_{N,n}
		\]
		and $f_{t_{n-1},t_n}(x_N)$ is the end position of the $N$th trajectory under the (numerically approximated) PUMA flow beginning at time $t_{n-1}$ and running until time $t_n$.
	\item[2c.] Determine the number of clones produced by each trajectory,
	\[
	c_{N,n} = \lfloor W_{N,n} + u_N \rfloor
	\]
	where $\lfloor \cdot\rfloor$ is the integer portion and $u_N$ are random variables generated from a uniform distribution on $[0,1]$.
	\item[2d.] The number of trajectories present after each iteration is given by,
	\[
	M_n = \sum_{N=1}^M c_{N,n}
	\]
	Clones are used as inputs into the next iteration of the algorithm. For large N, the normalizing	factor ensures the number of particles $N_n$ remains constant; however, in practice the number of particles fluctuates slightly on each iteration $n$. To ensure $N_n$ remains constant it is common to compute the difference $\Delta N_n = N_n-M$. If $\Delta N_i > 0$, then $\Delta N_i$ trajectories are randomly selected (without replacement) and killed. If $\Delta N_i<0$, then $\Delta N_i$ trajectories are randomly selected (with	replacement) and cloned. 
	\item[2e.] To ensure divergence, random uniform noise sampled from the interval $[-\varepsilon^{\circ}, +\varepsilon^{\circ}]$ with $\varepsilon = 10e-2$ is added to the clones.
	\end{itemize}
	\item[3.] The final set of positions $X_N = \phi\circ f_{t_{\lfloor \tau/\mathcal{T}\rfloor-1},t+\tau}(x_{N})$ tends to a new distribution as $N\rightarrow\infty$ exponentially tilted by the constant $C$.
\end{itemize}
\end{framed}

Since $\phi$ is maximized at $\gamma$, the set of end positions $X_N$ coming from GPA is the set with a higher probability of entering and remaining in the coherent set over the time interval $[t,t+\tau]$. Backwards reconstruction of the trajectories associated to the surviving end positions provides the set of initial conditions responsible for sending particles into the coherent set.

Given the $N=1,\dots,M$ trajectories $\{X_N(T)\}_{t\le T\le t+\tau} = \{\phi\circ f_{t_{n-1},t_n}(x_N)\}_{1\le n\le \lfloor \tau/\mathcal{T}\rfloor}$ run under GPA with exponential tilting function $V(x)$, it is shown in \cite{WB} that the expected value for any observable $F(\{X(T)\}_{t\le T\le t+\tau})$ of a trajectory $\{X(T)\}_{t\le T\le t+\tau} =  \{\phi\circ f_{t_{n-1},t_n}(x)\}_{1\le n\le \lfloor \tau/\mathcal{T}\rfloor}$ under the original distribution can be estimated as,
\begin{equation}\label{ex_obs}
E_0\big[F(\{X(T)\}_{t\le T\le t+\tau})\big] \sim \frac{1}{M}\sum_{N=1}^M F(\{X_N(T)\}_{t\le T\le t+\tau}) \times e^{V(X_N(t))-V(X_N(t+\tau))}\times \prod_{n=1}^{\lfloor \tau/\mathcal{T}\rfloor} Z_n
\end{equation}
This value is essentially the average value of the observable after the tilting effects on the importance sampled trajectory are removed. We first estimate the probability $p_{\phi}$ of a rare event occurring under our observable $F(x) = 1_{\{\phi\circ f_{t,\tau}(x)\ge \tilde{x}\}}(x)$, that is the probability of the observable $\phi$ is greater than some value $\tilde{x}$ at the final integration time $t+\tau$. We remark that $\phi$ is the exact observable whose distribution is tilted during genealogical particle analysis. For large enough values of starting particles $M$, the estimate $\hat{p}_{\phi}$ is normally distributed with mean $\mu = p_{\phi}$ and variance $\sigma_{\phi}^2$ \cite{WB}. The relative error is then estimated as $\text{RE} = \sigma_{\phi}/p_{\phi}$ or empirically by,
\begin{equation}\label{RE}
\text{RE} = \frac{\sqrt{\frac{1}{K}\sum_{J = 1}^K (\hat{p}_{\phi,J}-p_{\phi}})}{p_{\phi}}
\end{equation}
for $K$ runs of genealogical particle analysis where $p_{\phi}$ is estimated from a very long control run of the system.

In a similar way, if we are interested in estimating the rare event probability $p_{\mathcal{B}\rightarrow \mathcal{A}}$ of a trajectory starting in a region $\mathcal{B}\subset X$ and ending in a region $\mathcal{A}\subset A$ of the target coherent set $A$, we may define the observable, 
\begin{equation}\label{obs}
F(\{X_N(T)\}_{t\le T\le t+\tau}) = 1_{\{x_N \in \mathcal{B}, f_{t,\tau}(x_N)\in \mathcal{A}\}}\circ \phi^{-1}(\{X_N(T)\}_{t\le T\le t+\tau})
\end{equation}
as the indicator function of $x_N$ starting in region $\mathcal{B}$ at time $t$, flowing under $f_{t,\tau}$ until $t+\tau$, and ending in $\mathcal{A}$. Then the transition probability estimate $\hat{p}_{\mathcal{B}\rightarrow\mathcal{A}}$ is the value obtained from equation (\ref{ex_obs}) by plugging in equation (\ref{obs}). Once again, for large enough values of starting particles $M$, the estimate $\hat{p}_{\mathcal{B}\rightarrow\mathcal{A}}$ is normally distributed with mean $\mu = p_{\mathcal{B}\rightarrow\mathcal{A}}$ and variance $\sigma_{\mathcal{B}\rightarrow\mathcal{A}}^2$ \cite{WB}. The relative error is then estimated as in equation (\ref{RE}) by $\text{RE} = \sigma_{\mathcal{B}\rightarrow\mathcal{A}}/p_{\mathcal{B}\rightarrow\mathcal{A}}$.

GPA results in lower relative error for rare event probability estimates of $\phi\circ f_{t,\tau}$, or equivalently, probability estimates of trajectories ending in $\mathcal{A} := \{x:\phi(x) \ge \tilde{x}(C)\}$. We can then expect that these surviving trajectories also provide more accurate probability transition estimates from any starting region $\mathcal{B}\to\mathcal{A}$.
\section{An Application to the Portable University Model of the Atmosphere for Pollution Movement}

Studying the movement of particles along trajectories in the atmosphere is particularly interesting when tracking pollution movement. Eddies and other coherent structures in the atmosphere relate to slow mixing regions where pollution can be trapped for long periods of time. Often, studying pollution dispersion involves finding the trajectories of particles that are initially inside a coherent set and flow outward. In this section, we investigate the rare probability of certain regions in the atmosphere sending particles into the coherent set. Using the numerical techniques described above, we are able to more accurately estimate these tail probabilities compared to standard integration techniques.

\subsection{Locating a Target Coherent Set Over Europe}
We discretize the space $Y = [36^\circ, 70^\circ]\times[169^{\circ},205^{\circ}]$ (covering Europe) into a partition of $m = 4,896$ rectangles $\{B_1,\dots, B_m\}$ to form a uniform grid of $\frac{1}{2}^{\circ}\times\frac{1}{2}^{\circ}$ degree boxes. To calculate the transition probability matrix $P_{t,\tau}$, each box is filled with $N = 25$ uniformly distributed points $y_{i,l}$ for $i = 1,\dots, m$ and $l = 1,\dots, N$ and run under $f_{t,\tau}$ (over the interval $[t,t+\tau]$). $P_{t,\tau}$ is then estimated as,
\begin{equation}\label{TPM}
P_{t,\tau,i,j}\approx \frac{\#\{l: y_{i,l}\in B_i, f_{t,\tau}(y_{i,l})\in B_j\}}{N}.
\end{equation}
Calculation of $f_{t,\tau}(y)$ is done through a four-step Runge-Kutta procedure with a stepsize of $\frac{1}{10}$ of a day. Velocity field outputs from PUMA are given at 1-day intervals at 8,192 equally spaced grid points (64 latitude and 128 longitude) taken over whole domain $[-180^{\circ},180^{\circ}]\times[0^{\circ},360^{\circ}]$. We linearly interpolate over space and time to ensure a continuous flow.

\begin{figure}[h!]
	\includegraphics[width=0.6\textwidth, trim = 0 0 100 0]{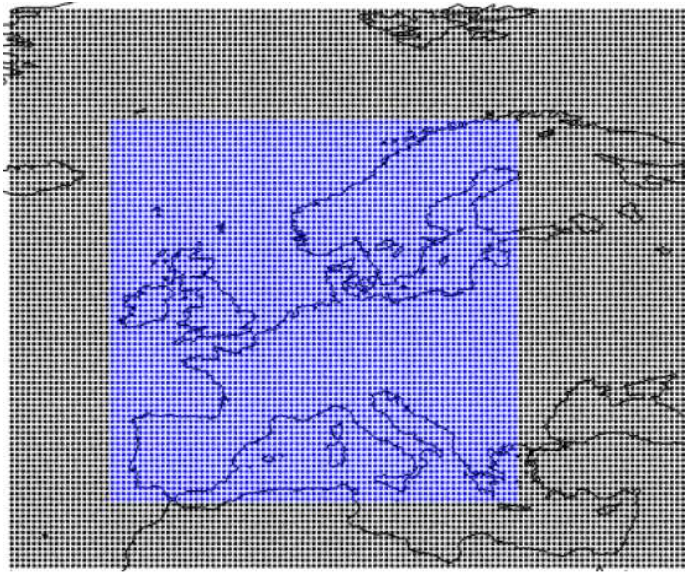}
	\caption{Grid of $B_i$ boxes where $Y = \cup_{i=1}^m B_i$ shown in blue. Initial values for importance sampling are taken uniformly over the region $X$ shown in black.\label{fig:domain}}
\end{figure}

We wish to estimate coherent sets that remain over Europe for $\tau = 4$ days. PUMA and many standard GCMs require time for the system to reach a state where the initial conditions of the model (temperature, pressure, wind velocity, etc.) provide meaningful and accurate time-series outputs. We choose a starting time of $t\ge 360$ days to ensure the model has reached this state. Furthermore, we choose our specific $t$ such that (at least) one eddy will pass through $Y$ over $[t,t+\tau]$. For our purposes $t = 360$.

To calculate the eigenvalues and eigenfunctions that will be used to form the partition of $Y$, we first calculate
\[
p_i = \frac{\text{Area of } B_i}{\text{Area of } Y}.
\]
Let $A = \cup_{i\in I} B_i$ with $I \subset \{1,\dots,m\}$ of indices, then from \cite[Prop. 6.4]{FD} we have
\[
\rho(A) \approx \frac{\sum_{i,j\in I} p_i P_{t,\tau,i,j}}{\sum_{i\in I} p_i}.
\]

Our measure $\mu$ is not (necessarily) invariant under $f_{t,\tau}$ because trajectories may begin in $Y$ but leave $Y$ through all boundaries. To ensure the matrix $P_{t,\tau}$ is stochastic and has $p$ as an exact fixed left eigenvector, we introduce a mixing tile \cite{LK} by adding a column at the $i = m+1$ position in $P_{t,\tau}$ with values $P_{t,\tau,i,m+1} = 1-\sum_{j=1}^m P_{t,\tau,i,j}$ and a row at the $j = m+1$ position with values $P_{t,\tau,m+1,j} = 1/(m+1)$.

Generally, $P_{t,\tau,i,j}$ is not reversible because,
\[
p_i P_{t,\tau,i,j}\ne p_j P_{t,\tau,j,i}.
\]
However, if we define the time-reversed quantity
\[
\hat{\rho}(A) = \frac{\sum_{i,j\in I} p_i \hat{P}_{t,\tau,i,j}}{\sum_{i\in I} p_i}
\]
where $\hat{P}_{t,\tau,i,j} = p_jP_{t,\tau,j,i}/p_i$ then in \cite{F} it is shown that $\rho(A) = \hat{\rho}(A)$. A major consequence of this relation is that the total cost function described in (\ref{graph2}) remains unchanged under time reversal and without loss of generality we may replace $P_{t,\tau,i,j}$ with the time reversible matrix with entries,
\[
R_{t,\tau,i,j} =\frac{1}{2} \bigg(P_{t,\tau,i,j}+\frac{p_jP_{t,\tau,j,i}}{p_i}\bigg)
\]

Following graph theory, we perform spectral clustering on the matrix $R_{t,\tau,i,j}$ to find the partitions $A_1,\dots,A_q$ which maximize (\ref{graph2}). We define the unnormalized graph Laplacian of the time reversible matrix $R_{t,\tau}$,
\[
L_{t,\tau} = D_{t,\tau}-R_{t,\tau}
\]
where $D_{t,\tau}$ is the diagonal matrix with entries $D_{t,\tau,i,i} = \sum_{i} R_{t,\tau,i,j}$. The matrix $L_{t,\tau}$ is symmetric and positive semi-definite with $0=\lambda_1\le\lambda_2\le\dots\le\lambda_m$. Moreover, solving for the optimal $q$ partition of $L_{t,\tau}$ is equivalent to solving a relaxation of the min-cut problem \cite{V}. We then perform standard $K$-means on the projection of $L_{t,\tau}$ onto the 1-dimensional subspace created from the second smallest eigenvalue of $L_{t,\tau}$. We refer the reader to the appendix \ref{KM} for a description of the algorithm.

\begin{figure}[h!]
	\includegraphics[width=0.6\textwidth,trim = 0 0 150 0]{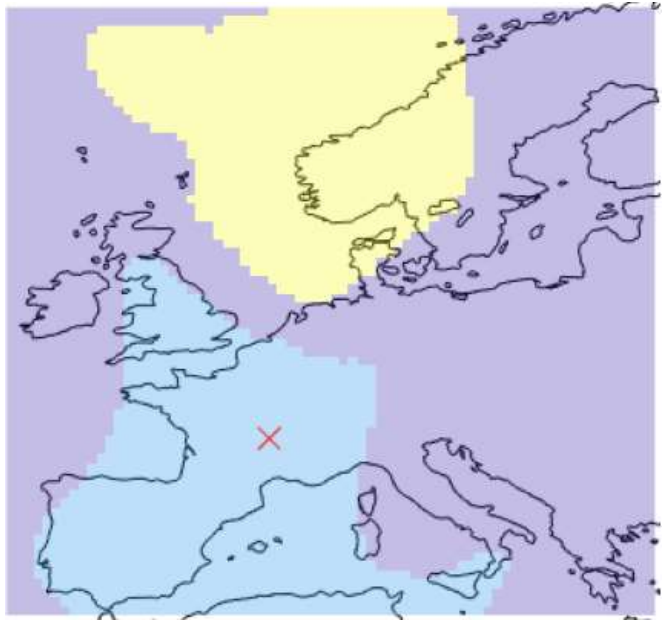}
	\caption{Two finite-time coherent sets under the flow $F$ over $[t,t+\tau] = [360,364]$. Regions are indicated by different colors. Center of the chosen target coherent set $A$ with $\rho(A) = 0.9771$ is marked with a red X.\label{fig:as}}
\end{figure}

We find two coherent sets over the domain $Y = [36^\circ, 70^\circ]\times[169^{\circ},205^{\circ}]$ representing the portion of the atmosphere over Europe. See Figure \ref{fig:as} for an illustration of these sets. A movie showing the graphic overlay of these sets with their time-dependent velocity fields and integrated path movement over $[t,t+\tau] = [360, 364]$ is provided in the Supplementary Material. We remark that the chosen target coherent set is the result of two eddy interactions (illustrated in the movie).

\begin{figure}[h!]
	\includegraphics[width=0.6\textwidth, trim = 110 200 100 200]{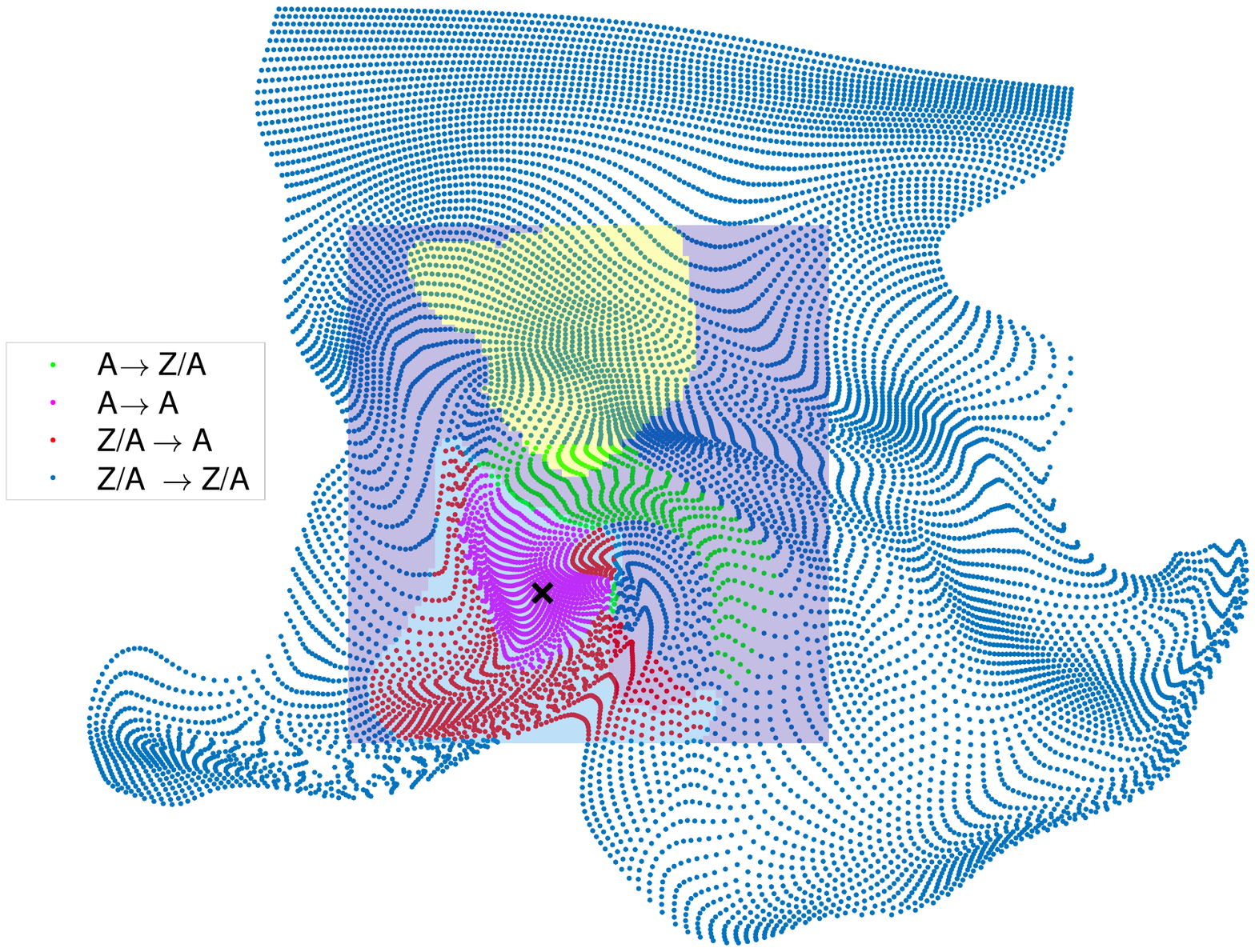}
	\caption{End position of trajectories under the flow. Different colors illustrate regions of mixing: (green) starting from the coherent set $A$ and ending in the external space $Z/A$, (purple) starting in $A$ and ending in $A$, (red) starting in $Z/A$ and ending in $A$ and (blue) starting in $Z/A$ and ending in $Z/A$. The center of the coherent set is marked with an X.\label{fig:mix}}
\end{figure}

Our main motivation for performing this analysis is to better estimate the probability of a starting region sending trajectories into the target coherent set. Since it is rare under the flow $f_{t,\tau}$ for trajectories to enter the set, a large number of initial conditions are required to estimate this probability through standard integration techniques. Figure \ref{fig:mix} shows the movement of a uniform grid of starting particles under the map without importance sampling. We now illustrate that the backward reconstructed surviving trajectories of genealogical particle analysis can provide more accurate estimates of these probabilities under an equivalent computational cost.

\begin{figure}[h!]
	\includegraphics[width=0.6\textwidth, trim = 100 0 100 0]{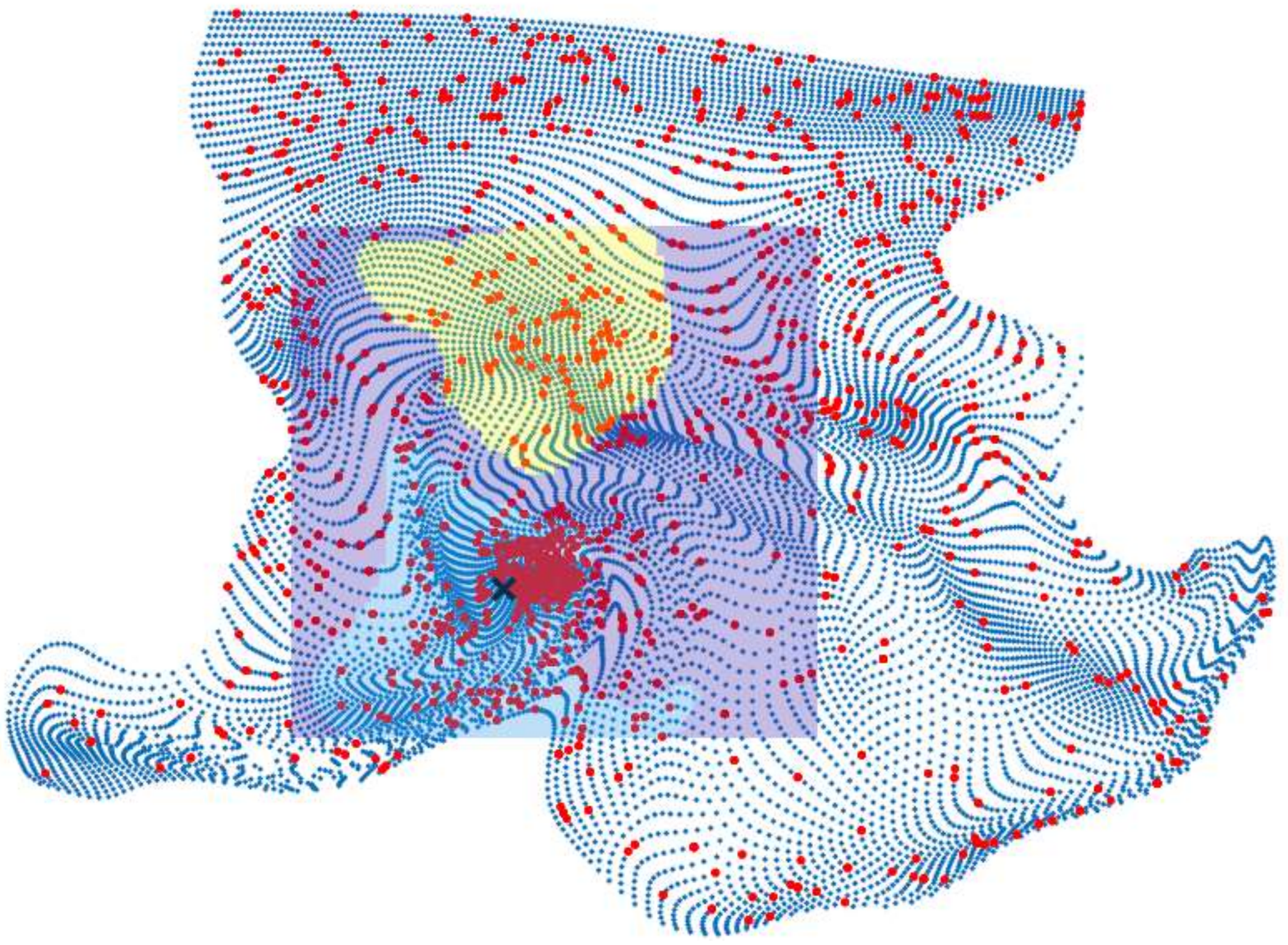}
	\caption{End position of trajectories (blue) under the flow and (red) after genealogical particle analysis. Note that most surviving trajectories end near the center of the coherent set (marked with an X).\label{fig:gpa_end}}
\end{figure}

We run GPA on $M = 200$ initial particles uniformly sampled over the domain $X=[30^{\circ}, 80^{\circ}]\times[160^{\circ},220^{\circ}]$ and varying values of $C$ ($C=0$ is a control value) with sampling time $\mathcal{T} = 0.01$ over the time interval $[t,t+\tau]$. For pragmatic reasons, we add a constant value $D$ to the observable so that $\phi(x) = -\log(d(x,\gamma))+D\ge 0$. This shift by $D$ ensures that negative values in the exponent of the weight function are the result of a true decrease of the observable value from the previous step. Figure \ref{fig:gpa_end} shows an example of the end \textit{location} distribution of surviving trajectories after genealogical particle analysis. As expected, most surviving trajectories are located near the center of the coherent set. Since GPA only increases the \textit{probability} of observing trajectories that end near the center, we still expect a (smaller) portion of end positions across the full domain.

\begin{figure}[h!]
	\includegraphics[width=0.6\textwidth, trim = 80 180 80 180]{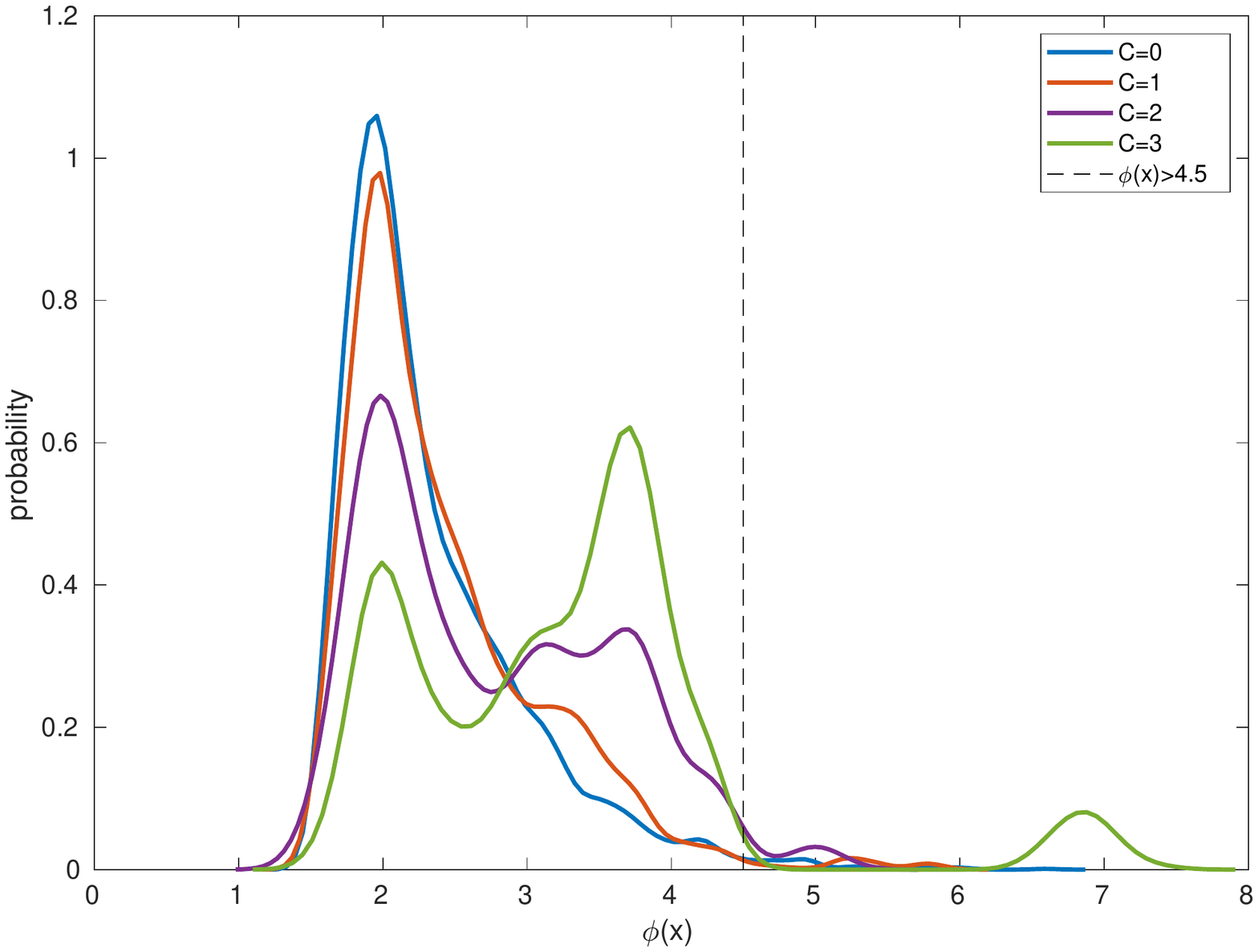}
	\caption{Exponentially tilted distributions for different $C$ values under GPA. Densities are estimated with a normal kernel. $C=0$ corresponds to the original distribution.\label{fig:dist}}
\end{figure}

\begin{figure}[h!]
	\includegraphics[width=0.8\textwidth]{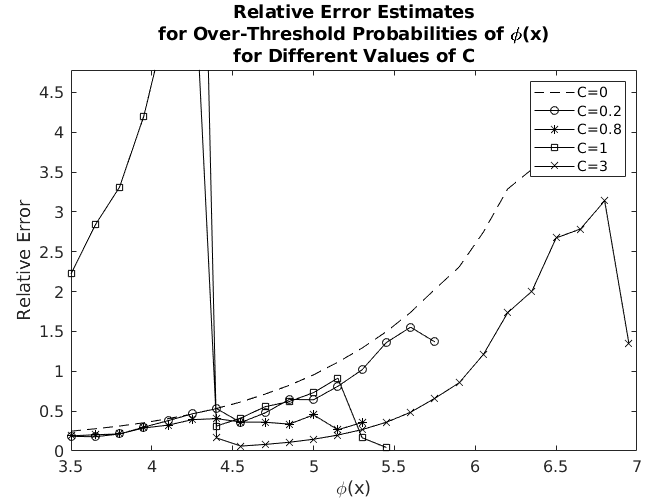}
	\caption{Relative error on the over threshold probability $p_{\phi}$ for different $C$ values under GPA.\label{fig:RE_phi}}
\end{figure}

We estimate the density functions (using a normal kernel) for $M = 200$ of the end positions at time $t+\tau$ after GPA sampling for each value of $C$. As $C$ increases, the probability of sampling larger values of $\phi$ increases. We refer to Figure \ref{fig:dist} for plots of these densities. Probability estimates $p_{\phi}$ are calculated using equation (\ref{obs}) with the relative error estimated by equation (\ref{RE}) with $K=20$ runs of genealogical particle analysis for each value of $C$. Results are shown in Figure \ref{fig:RE_phi}.

We divide the domain $X=[30^{\circ}, 80^{\circ}]\times[160^{\circ},220^{\circ}]$ into $5^{\circ}\times 5^{\circ}$ boxes, $\mathcal{B}$, and estimate probability transitions $\hat{p}_{\mathcal{B}\rightarrow\mathcal{A}}$ to the (circular) region $\mathcal{A}$ with center $\gamma$ so that $\mathcal{A}:=\{x:\phi(x)\ge \tilde{x}(C)\}$ where $\tilde{x}(C)$ is the value of $\phi(x)$ such that the relative error given by equation (\ref{RE}) with $p_{\phi}$ is smaller than that of a brute force estimate of equivalent computational effort. We choose $\phi(x)\ge \tilde{x}(C) = 4.5$ conservatively, but note that smaller values of $\phi(x)$ (corresponding to a larger region $\mathcal{A}$) can be estimated with smaller or combined choices of $C$. We choose the same value $\tilde{x}(C)$ for all choices of $C$ in this analysis for cross comparison. We then calculate the transition probability estimates $\hat{p}_{\mathcal{B}\rightarrow\mathcal{A}}$ of every $5^{\circ}\times 5^{\circ}$ box, $\mathcal{B}$, given by equation (\ref{obs}) using (a) the surviving trajectories of GPA and (b) the brute force simulation of equal computational effort. This comparison is done by calculating the relative error estimate on the transition probabilities given by equation (\ref{RE}) with $p_{\mathcal{B}\rightarrow\mathcal{A}}$. All long control runs are estimated with $M = 80,000$ starting particles.

\begin{figure}[h!]
	\includegraphics[width=0.8\textwidth]{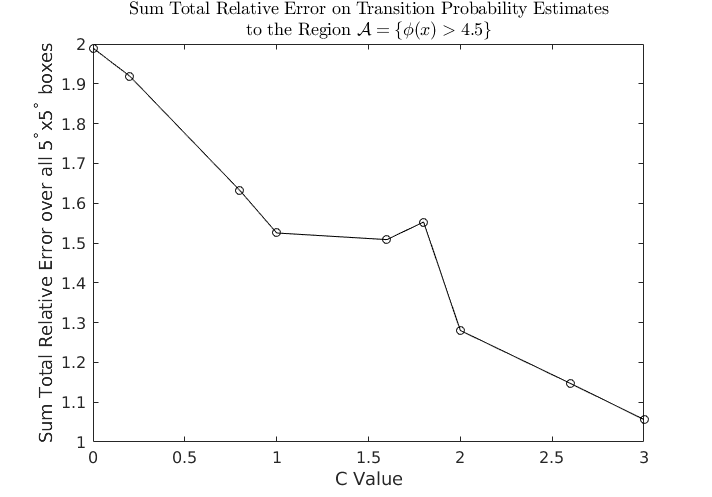}
	\caption{Sum total error on the transition probability $p_{\mathcal{B}\rightarrow \mathcal{A}}$ over all boxes $\mathcal{B}$ for different values of $C$ and $M = 200$ starting particles. $C=0$ corresponds to the brute force estimate of the same computational effort. \label{fig:RE_TRNS}}
\end{figure}

\begin{figure}[h!]
	\includegraphics[width=\textwidth, trim = 0 180 0 180]{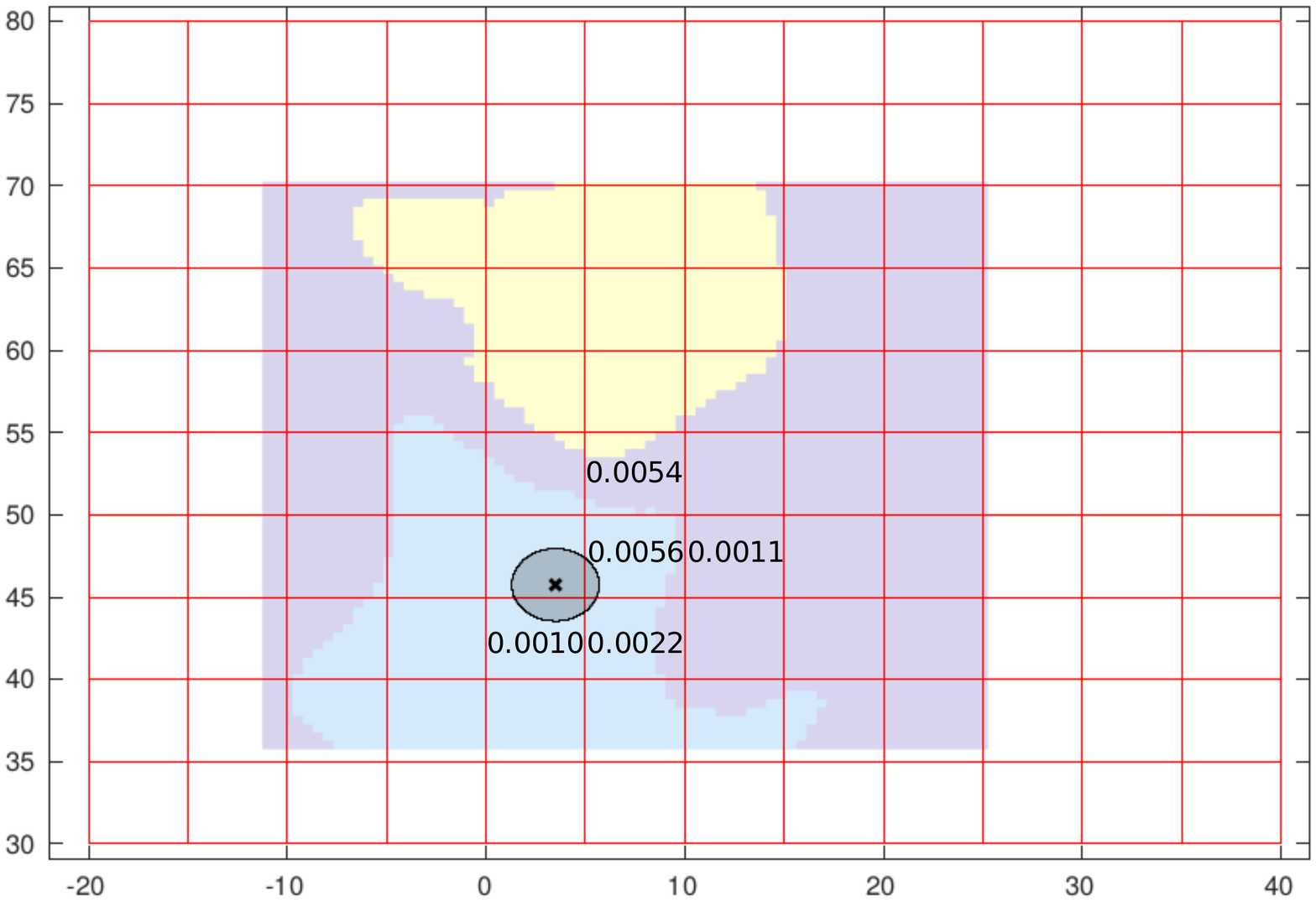}
	\caption{Brute force and GPA ($C=3$) difference in total error of transition probability estimates $\hat{p}_{\mathcal{B}\rightarrow\mathcal{A}}$ for every $5^{\circ}\times 5^{\circ}$ box $\mathcal{B}$ into $\mathcal{A}$ (represented by a black circle). Blank squares correspond to $0$ difference. All values are positive indicating transition estimates using trajectories from GPA with $C=3$ are as good or more accurate than those of brute force for all $\mathcal{B}$. \label{fig:RE_GRID}}
\end{figure}

We calculate the relative error on the transition probability for each starting box $\mathcal{B}$ on the grid. We show that for the region $\mathcal{A} = \{x:\phi(x)\ge 4.5\}$, transition probabilities from $\mathcal{B}\rightarrow A$ have relative error that decreases for increasing $C$. Figure \ref{fig:RE_TRNS} shows the sum total of error over all boxes $\mathcal{B}$ decreases. Figure \ref{fig:RE_GRID} compares the brute force relative error with those of $C=3$ with the same computational cost ($M=200$) where we find lower error over every box in the grid. From these results, we conclude that surviving trajectories of GPA can provide more accurate rare event transition probabilities where a rare event in this setting is defined as a trajectory entering the subset $\mathcal{A}$ of a target coherent set $A$. 

\section{A Brief Discussion on this Method as an Application for Storm Tracking}
We have shown in the previous example that the methods outlined in this paper can be used to track the collection of particles (or pollution) in the atmosphere to a fixed coherent set that is formed by the background movement of counter-rotational eddies through the space. Now we consider a finite-time coherent set formed from a single eddy. Storm systems, such as hurricanes, have properties similar to that of a coherent set so we can use these naturally occurring atmospheric eddies as a foundational model for storm movement.

We investigate the coherent set as a function of a shorter time step and use importance sampling to find its most likely path. Our hypothesis, which we will test numerically using the method described in this paper, is that the set of regions with the highest transition probability of initial values ending near the center of a finite-time coherent set in the current step, will provide the most likely direction of the center of the finite-time coherent set in the next step. Each step in the path is determined by a transition probability matrix, built over a small time interval on which the coherent set is defined, with states given by a spacial grid. Transitions are taken as the probability of trajectories starting in a region and ending near the center of the coherent set. Estimates of these probabilities are found by using genealogical particle analysis to enrich this set of trajectories and obtain more accurate transition estimates. As a rule, all notation in this section is carried over from the previous example.

We numerically approximate the flow built from the same northern and eastern velocity field outputs of PUMA by the Runge-Kutta method described previously. Next, we find a target coherent set in the region $Y\subset X$ over $J = 3$ non-overlapping, consecutive time windows of length equal to $\tau = 1$ day, $[t+J,t+J+\tau] = [360+J, 360+J+1]$. The result is a time-dependent coherent set found over three discrete time intervals; one set is found for each time interval. The length of the chosen time intervals is relative to the movement speed of the coherent set. Time windows of a shorter length do not show a significant amount of movement of the coherent set while time windows of a longer length produce overlapping eddies resulting in coherent sets of a different form. 

\begin{figure}
\includegraphics*[width = \textwidth]{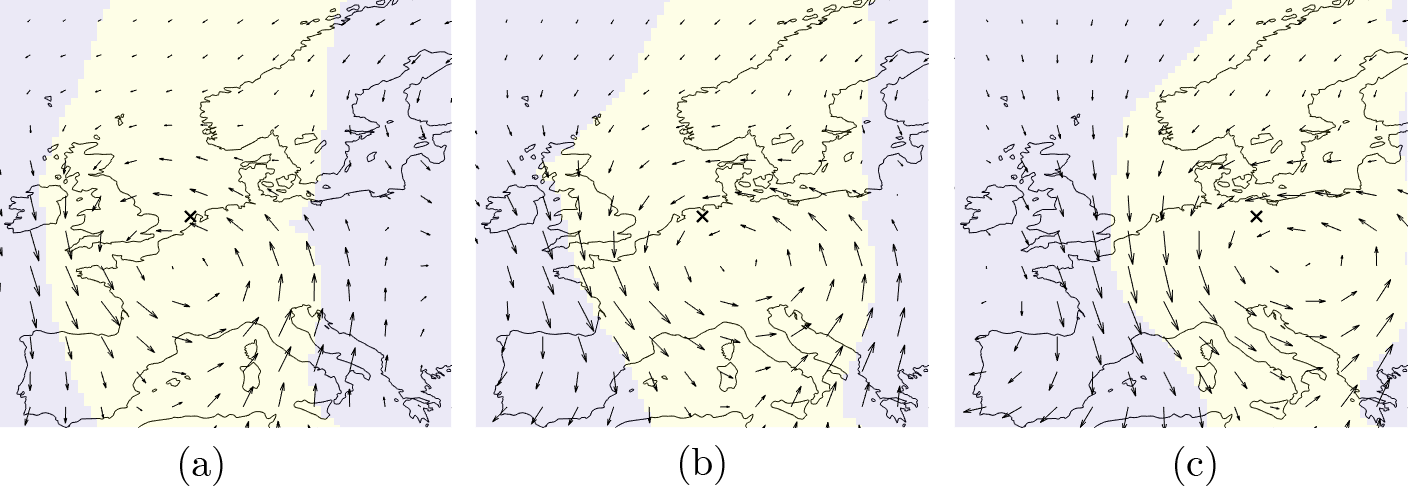}
\caption{Coherent sets found from the transfer operator method of $J = 3$ non-overlapping, consecutive time windows of length 1 day. Time intervals $[t+J,t+J+\tau]$ = (a) $[360,361]$ (b) $[361, 362]$ (c) $[362,363]$. Quivers indicate the velocity field.}
\end{figure}

For each $J$, we run genealogical particle analysis using tilting value $C=3$ on the set of uniformly distributed particles over $X$ with starting time $t+J$ and termination time $t+J+1$. Recall that GPA returns a set of trajectories that behave as though they come from the exponentially tilted distribution where there is a higher likelihood of obtaining larger values of the observable $\phi(x) = -\log d(x,\gamma(A(J)))$ where $\gamma(A(J))$ is the midpoint (center) of the $J$th corresponding coherent set. Hence, the outcome is the set of trajectories most likely to end near $\gamma(A(J))$. The resampling time is taken at $\mathcal{T}= 0.1$ with sampling times $t_n = \mathcal{T}n$, $n = 1,\dots,\lfloor \tau/\mathcal{T}\rfloor = 10$. Backwards reconstruction of surviving trajectories is then used to determine the set of initial points which are most likely to end near  $\gamma(A(J))$.

\begin{figure}
\includegraphics[width=\textwidth]{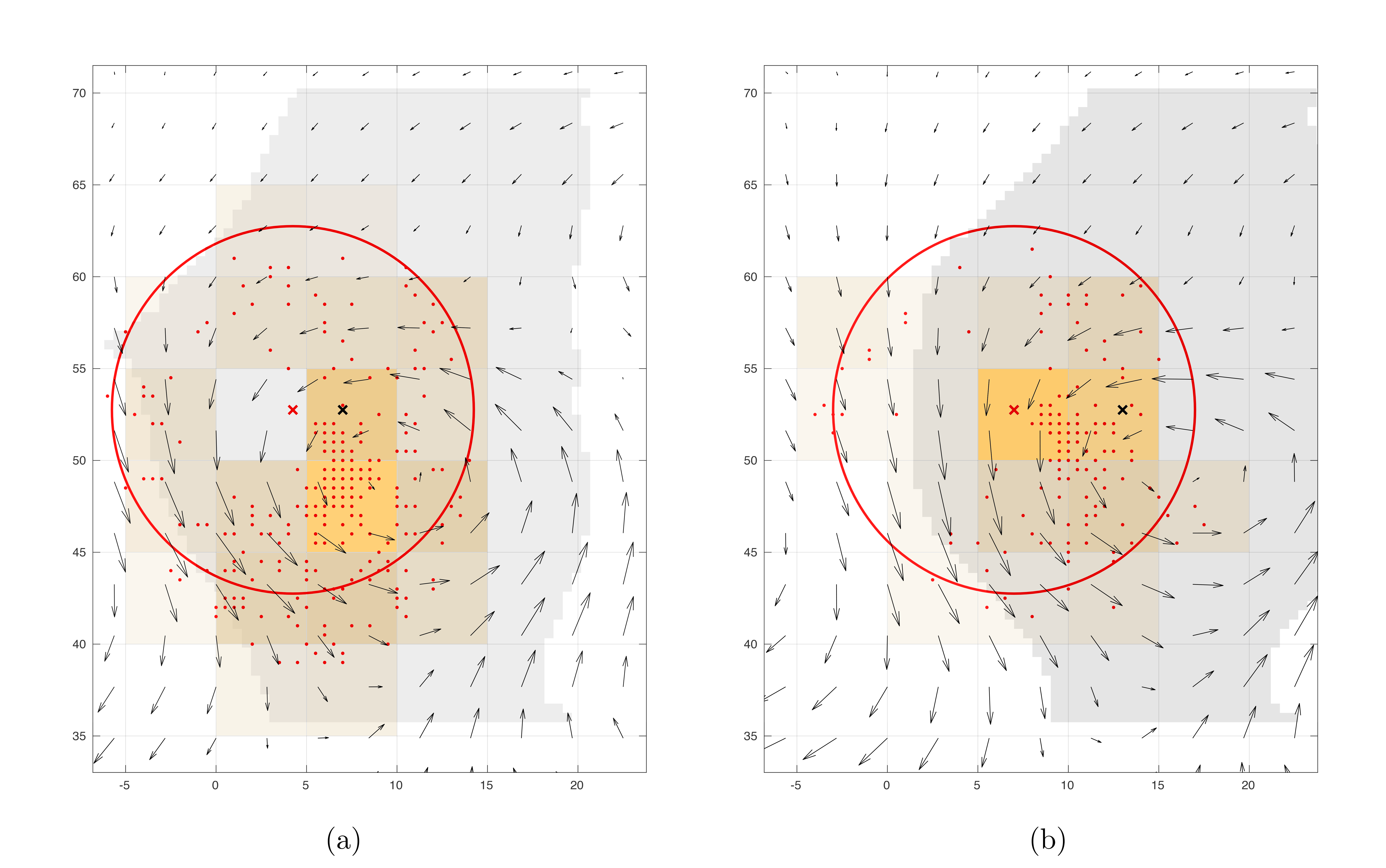}
\caption{Movement prediction of the coherent set. The center for step $J$ is marked with a red X and the corresponding initial positions of the end surviving trajectories within $B(\gamma(A(J)),r)$ are indicated by red points. The $J+1$ coherent set is highlighted in gray, its corresponding velocity field is represented by black quivers and center is marked with a black X. (a) $J = 1$ and (b) $J = 2$. The movement direction probability is taken as the proportion of (surviving) initial positions in the regions marked by the grid. These probabilities are highlighted in orange with darker values indicating a higher probability. \label{fig:storm_move}}
\end{figure}

In this example, the set of possible starting regions is defined after GPA as the set of $5^{\circ}\times 5^{\circ}$ boxes covering all of $X$. For each $J$, we have an associated region $E_{J}\subset X$ corresponding to the starting region that has the highest proportion of initial points from surviving trajectories (over all starting regions). Since $E_{J}$ has the highest probability of sending trajectories near $\gamma(A(J))$ at time $t+J+1$, this region should provide us with the movement direction of the coherent set $A(J)\rightarrow A(J+1)$ (and its corresponding center $\gamma(A(J))\rightarrow\gamma(A(J+1))$) defined over $[t+J+1,t+J+2]$. Using each of the $J=3$ invariant sets found previously from the PUMA flow approximation, we illustrate in Figure \ref{fig:storm_move} that $E_J$ can provide some reasonable indication of movement direction for the coherent set in the next time step.

\section{Conclusion}

Coherent sets in the atmosphere are physically interesting because they relate to single eddies and eddy interactions. In this investigation, we look at coherent structures in the atmosphere represented numerically by the 2-dimensional flow of the Portable University Model of the Atmosphere. We use a modified set of tools revolving around a well-studied transfer operator approach to estimate these regions. In particular, we approximate the Perron-Frobenius operator by the transition probability matrix for the flow over the European subregion and use spectral $K$-means clustering to find coherent sets only located over Europe. 

It can be seen for longer time intervals that the coherent sets are formed by multiple eddy interactions such as particles trapped between the spin of two eddies; whereas shorter time intervals have coherent sets corresponding to a single eddy. For the former, one can ask questions about the regional origin of trajectories ending inside the coherent set and their most probable paths. For the latter, one can ask questions about the path of such an eddy in the space. To study these trajectories, we employ a well-known importance sampling algorithm, called genealogical particle analysis, not used in this context to-date.

Current literature has focused on using importance sampling methods to decrease the relative error of an estimated rare event probability by forcing rare events to occur more frequently. We show that these methods can also provide useful information on the set of trajectories likely to end in an extreme event. For the interest of this study, we have introduced an observable that defines the extreme event as being near the center of an coherent set. In this setting, we show that the surviving trajectories obtained from importance sampling can provide information on probable paths and initial regions of trajectories that end in an coherent set under an atmospheric flow. We complete our investigation by motivating and illustrating some important examples where information about trajectory movement toward the center of a coherent set in the atmosphere is useful and physically relevant: origin of pollution and storm movement. We provide numerical evidence that these surviving trajectories give more accurate probability transition estimates compared to standard integration techniques.

In future work we plan to apply these techniques to real hurricane data where the fixed point $\gamma$ may be taken as some point outside of the finite-time coherent set. The outcome of importance sampling would then give us the probability of a hurricane moving over a given region. It would also be interesting to consider importance sampling methods for the sequence of maxima $M_n = \max\{X_1,\dots,X_N\}$ where $X_N = \phi\circ f(x_N)$. This would limit the set of original distributions to the family of generalized extreme value functions and possibly provide a new way of using the Hausdorff distance in the definition of $\phi(x) = -\log(d(x,\gamma))$. Furthermore, a complete shift of the generalized extreme value distribution under exponential tilting would result in a higher density around $\gamma$ and less uniformly distributed points about the whole space.

\clearpage
\section*{Supplementary Material}
See supplementary material for a movie illustrating the coherent sets found over the whole interval $[360,364]$ (4 days), time-dependent velocity fields and integrated paths taken at $1/10^{\text{th}}$ day time steps.

\section*{Acknowledgments}
Special thanks to Matthew Nicol for his expertise and advice on the foundations of this paper. Thanks to Frank Lunkeit for helpful discussions and information on the Portable University Model of the Atmosphere.

\section*{Data Availability}
All data used in this paper was generated from the Portable University Model of the Atmosphere. This atmospheric model is freely available for download at the Universit\"{a}t Hamburg website for Planet Simulator \cite{LBFJKLS}.

\appendix
\section{Spectral Clustering with $K$-Means}\label{KM}
\begin{itemize}
	\item[1.] Form the unnormalized Laplacian of the matrix $R_{t,\tau}$,
	\[
	L_{t,\tau} = D_{t,\tau}-R_{t,\tau}
	\]
	where $D_{t,\tau}$ is the diagonal matrix with entries $D_{t,\tau,i,i} = \sum_{i=1}^m R_{t,\tau,i,j}$.
	\item[2.] Choose the first $\ell$ eigenvalues $\lambda_1,\dots,\lambda_{\ell}$ and corresponding eigenvectors $v_1,\dots,v_{\ell}$ of $L_{t,\tau}$. 
	\item[3.] Form a subspace made of $S_\ell := \text{sp}\{v_1,\dots,v_{\ell}\}$ and project the $m$-dimensional row vectors of $L_{t,\tau}$ onto $S_\ell$.
	\item[4.] Run $K$-means on the projected $\ell$-dimensional row vectors with a predetermined $K$ value. $K$ essentially tells the algorithm how many coherent sets are expected. 	
The standard $K$-means algorithm for a set of $m$ nodes $n$ represented by $m$ vectors in $\mathbb{R}^H$ is given by,
\begin{itemize}
\item[4a.] Start with $K$ random partitions $P_j$ of the space $\mathbb{R}^H$.
\item[4b.] Compute the centroids (means) of these partitions as $C_j = \sum_{n(\ell)\in P_j} n(\ell)/\text{card}(P_j)$ where $C_j\in \mathbb{R}^H$.
\item[4c.] \textbf{Assign} $n(\ell)$ to the partition $P_j$ with the minimum (squared) euclidean distance between $n(\ell)$ and $C_j$.
\item[4d.] \textbf{Update} the algorithm by recalculating the centroids (means) of $P_j$.
\end{itemize}
The algorithm continues by repeating steps 4c and 4d until the assignments no longer change. This is equivalent to finding the steady state of the objective function given by,
\[
\min \sum_{j=1}^K \sum_{\ell}^m ||n(\ell)-C(j)||_{\mathbb{R}^H},
\]
the minimum sum of the (squared) euclidean distances between each node and its assigned centroid.
\end{itemize}

\end{document}